\documentclass[aps,preprint]{revtex4}%
\usepackage{amsfonts}
\usepackage{amsmath}
\usepackage{amssymb}
\usepackage{appendix}
\usepackage[T1]{fontenc}
\usepackage{graphicx}%
\ifx\pdfoutput\relax\let\pdfoutput=\undefined\fi
\newcount\msipdfoutput
\ifx\pdfoutput\undefined\else
\ifcase\pdfoutput\else
\ifx\paperwidth\undefined\else
\ifdim\paperheight=0pt\relax\else\pdfpageheight\paperheight\fi
\ifdim\paperwidth=0pt\relax\else\pdfpagewidth\paperwidth\fi
\fi\fi\fi
\begin{document}
\preprint{ }
\title[Light-matter interactions]{Fundamental formulation of light-matter interactions revisited}
\author{H. R. Reiss}
\affiliation{Max Born Institute, Berlin, Germany}
\affiliation{American University, Washington, DC, USA}
\email{reiss@american.edu}


\begin{abstract}
The basic physics disciplines of Maxwell's electrodynamics and Newton's
mechanics have been thoroughly tested in the laboratory, but they can
nevertheless also support nonphysical solutions. The unphysical nature of some
dynamical predictions is demonstrated by the violation of symmetry principles.
Symmetries are fundamental in physics since they establish conservation
principles. The procedures explored here involve gauge transformations that
alter basic symmetries, and these alterations are possible because gauge
transformations are not necessarily unitary despite the widespread assumption
that they are. That gauge transformations can change the fundamental physical
meaning of a problem despite the preservation of electric and magnetic fields
is a universal proof that potentials are more basic than fields. These
conclusions go to the heart of physics. Problems are not evident when fields
are perturbatively weak, but the properties demonstrated here can be critical
in strong-field physics where the electromagnetic potential becomes the
dominant influence in interactions with matter.

\end{abstract}
\date[12 October 2019]{}
\maketitle

\section{Introduction}

The unfamiliar properties of electromagnetism to be described here can be
overlooked when the electromagnetic field is no more than a perturbative
influence in physical processes. However, when the electromagnetic field is
the dominant influence, then these properties become profoundly important. The
ever-expanding use of powerful lasers imparts a fundamental significance to
these unfamiliar properties.

Current beliefs about electromagnetism that are challenged here come from the
demonstrations that: electromagnetic potentials convey more physical
information than electric and magnetic fields; reliance on electric and
magnetic fields can introduce basic errors; gauge transformations alter the
properties of a physical system; gauge transformations are not unitary;
concepts such as the adiabaticity property in laser phenomena are false and
wasteful; a proposed nondipole correction is unphysical; and predictions that
follow from the solutions of Maxwell's equations can be unphysical. From these
results, it is a corollary that Newton's mechanics can also support unphysical solutions.

An ancillary matter is the objection to the widespread use of an intensity
parameter that lacks Lorentz invariance, but is held to be descriptive of
otherwise covariantly-described phenomena.

It is emphasized that neglect of basic electrodynamic principles in
applications to strong-field laser processes has caused important hindrances
to the development of the discipline. These hindrances continue, and can lead
to needless delays in the development of this large and expanding field of study.

\section{Gauge transformations alter physical properties}

The fact that gauge transformations can fundamentally alter the physical
identity of a system is evident even in the elementary problem of an electron
immersed in a uniform constant electric field $\mathbf{E}_{0}$.

A possible set of potentials to describe the field is%
\begin{equation}
\phi=-\mathbf{r\cdot E}_{0},\quad\mathbf{A}=0. \label{A}%
\end{equation}
The Lagrangian that describes the electron in the field is independent of
time. By Noether's Theorem \cite{noether}, this means that energy is
conserved. Another possible gauge for the description of the constant field is%
\begin{equation}
\phi=0,\quad\mathbf{A}=-c\mathbf{E}_{0}t. \label{B}%
\end{equation}
The Lagrangian for an electron with the field described by Eq. (\ref{B}) has
time dependence, but is independent of the spatial coordinate $\mathbf{r}$. In
this case, energy is not conserved but momentum is conserved.

The potentials (\ref{A}) and (\ref{B}) have different symmetries, and
represent different physical situations. These differences are produced by the
gauge transformation.

An important case that possesses only one gauge that satisfies all relevant
symmetries was examined in Ref. \cite{hrgge}. The electromagnetic field
examined is a plane-wave field, such\ as that of a laser beam. The symmetry
that is present in that case is the propagation property, which requires that
the field can depend on the spacetime 4-vector $x^{\mu}$ only as a scalar
product with the propagation 4-vector $k^{\mu}$:
\begin{equation}
\varphi\equiv k^{\mu}x_{\mu}=\omega t-\mathbf{k\cdot r,} \label{C}%
\end{equation}
where $\omega$ is the field frequency and $\mathbf{k}$ is the propagation
3-vector. When a scalar potential $\phi$ such as that from a Coulomb potential
is also present, then the sole possible gauge satisfying the necessary
symmetry is the radiation gauge (also called Coulomb gauge), where the
3-vector component $\mathbf{A}$ is descriptive of the plane-wave field and the
scalar potential $\phi$ describes the binding potential, so that the total
4-potential is%
\begin{equation}
A^{\mu}:\left(  \phi^{scalar},\mathbf{A}^{planewave}\right)  . \label{D}%
\end{equation}

\section{Gauge transformations are not necessarily unitary}

The starting point here is the property known as form invariance, where the
Schr\"{o}dinger equation has the same form when expressed in terms of the
gauge-transformed potentials as it does in the original gauge. See, for
example, Ref. \cite{CT}. Form invariance under a gauge transformation
generated by the operator $U$ can be written as%
\begin{equation}
\widetilde{H}-i\hslash\partial_{t}=U\left(  H-i\hslash\partial_{t}\right)
U^{-1}, \label{E}%
\end{equation}
where $\widetilde{H}$ is the transformed Hamiltonian. This gives the
gauge-transformed Hamiltonian%
\begin{equation}
\widetilde{H}=UHU^{-1}-i\hslash U\left(  \partial_{t}U^{-1}\right)  .
\label{F}%
\end{equation}
This means that the gauge transformation cannot be a unitary transformation if
$U$ is time-dependent.\newline

For laser-related problems, the time dependence of the field imparts time
dependence to any gauge transformation employed. Such transformations are not,
in general, unitary.

\section{Electromagnetic potentials are more fundamental than electric and
magnetic fields}

The primacy of potentials over fields was first established by the
Aharonov-Bohm effect \cite{es,ab}. This relates to a specific example: the
deflection of an electron beam as it moves in the field-free region around a
solenoid. It is the potential that causes the deflection, since there is a
potential but no field outside the solenoid. That quantum result stood for
many years as the sole example of the fundamental role of electromagnetic
potentials. A more general case is the demonstration \cite{hrgge} that there
exists an unphysical solution of the Maxwell equations for a plane-wave field
propagating in the vacuum. This has consequences that are both quantum and classical.

Furthermore, as shown in the following Section, when a solution of the Maxwell
equations is unphysical, then the properties of the potentials are necessary
to distinguish physical from unphysical solutions. This is a universal proof
that potentials are more fundamental than fields.

\section{Solutions of Maxwell equations are not necessarily physical}

A single unphysical solution of Maxwell's equations is sufficient to
demonstrate that such unphysical solutions can exist. The example selected
here is significant since it has been proposed or employed for practical
laser-induced processes.

The symmetry condition that applies to all plane-wave fields, such as laser
fields, comes from the Einstein Principle \cite{einstein} that the speed of
light in vacuum is the same in all inertial frames of reference. This was
referred to above as the propagation property. Its mathematical statement is
that the spacetime 4-vector $x^{\mu}$ can occur only as the scalar product
$\varphi$ defined in Eq. (\ref{C}); that is, the vector potential describing a
plane wave must have the form $A^{\mu}\left(  \varphi\right)  $.

A gauge transformation of the electromagnetic field is generated by the
function $\Lambda$:%
\begin{equation}
A^{\mu}\rightarrow\widetilde{A}^{\mu}=A^{\mu}+\partial^{\mu}\Lambda. \label{G}%
\end{equation}
The only constraints on $\Lambda$ are that it be a scalar function and that it
satisfies the homogeneous wave equation%
\begin{equation}
\partial^{\mu}\partial_{\mu}\Lambda=0. \label{H}%
\end{equation}
This is sufficient to preserve the electric and magnetic fields. If $A^{\mu}$
satisfies the Lorenz condition $\partial^{\mu}A_{\mu}=0$, the same will be
true of $\widetilde{A}^{\mu}$. Now consider the generating function
\cite{hr79}
\begin{equation}
\Lambda=-A^{\mu}x_{\mu}, \label{I}%
\end{equation}
which leads to the gauge-transformed potential%
\begin{equation}
\widetilde{A}^{\mu}=-k^{\mu}\left(  x^{\nu}A_{\nu}^{\prime}\right)  ,
\label{J}%
\end{equation}
where $A_{\nu}^{\prime}$ is the total derivative of $A_{\nu}$ with respect to
$\varphi$: $A_{\nu}^{\prime}=\left(  d/d\varphi\right)  A_{\nu}$. Equation
(\ref{J}) takes a familiar form if the initial gauge for $A^{\mu}$ is the
radiation gauge. A pure plane-wave field is described in the radiation gauge
by the 4-vector%
\begin{equation}
A^{\mu}\left(  \varphi\right)  :\left(  0,\mathbf{A}\left(  \varphi\right)
\right)  . \label{K}%
\end{equation}
The gauge-transformed 4-vector is then%
\begin{equation}
\widetilde{A}^{\mu}=-\widehat{k}^{\mu}\mathbf{r\cdot E}\left(  \varphi\right)
,\quad\widehat{k}^{\mu}\equiv\frac{k^{\mu}}{\omega/c}, \label{L}%
\end{equation}
where $\widehat{k}^{\mu}$ is the unit propagation 4-vector that lies on the
light cone.

The form (\ref{L}) resembles the dipole-approximation scalar potential
$\mathbf{r\cdot E}\left(  t\right)  $ that is so ubiquitous in length-gauge
Atomic, Molecular, and Optical (AMO) physics. This is the reason why it was
examined in Ref. \cite{hr79} in an attempt to provide a rigorous basis for the
Keldysh approximation \cite{keldysh} of strong-field atomic physics. It was
rejected in Ref. \cite{hr79} on multiple grounds, the most obvious of which is
that it violates the Einstein Principle. The violation is evident in Eq.
(\ref{J}) from the appearance of $x^{\nu}$ in isolation from the propagation
4-vector, and the presence of the 3-vector $\mathbf{r}$ in Eq. (\ref{L}) that
requires an origin for a fixed spatial coordinate system that is contrary to
the nature of a freely propagating plane-wave field. Nevertheless, the fields
are preserved by the gauge transformation (\ref{I}), and so are the Lorenz
condition $\partial^{\mu}A_{\mu}=0$ and the transversality condition $k^{\mu
}A_{\mu}=0$ \cite{hr79}.

The 4-potential in Eq. (\ref{J}) or (\ref{L}) has the curious feature that it
lies on the light cone. A plane-wave field is described by a spacelike
4-potential, not one that is lightlike. Furthermore, a fundamental property of
a charged particle in interaction with a plane-wave field is the ponderomotive
energy \cite{hrdiss,hr62,hrup} $U_{p}$, which is proportional to $A^{\mu
}A_{\mu}$. However, since $k^{\mu}$ is self-orthogonal,
\begin{equation}
k^{\mu}k_{\mu}=0, \label{M}%
\end{equation}
the $\widetilde{A}^{\mu}$ of Eq. (\ref{J}) or (\ref{L}) predicts a zero
ponderomotive energy for any charged particle.

For all of these reasons, Eqs. (\ref{J}) and (\ref{L}) are unphysical.
Nevertheless, they are arrived at by a valid gauge transformation from a
proper plane-wave 4-potential, meaning that they predict the same electric and
magnetic fields, and hence they satisfy the same Maxwell equations, since the
Maxwell equations depend only on the fields, not on the potentials. This is
proof that Maxwell's equations can support unphysical solutions.

Equations (\ref{J}) and (\ref{L}) were first proposed and discussed in Ref.
\cite{hr79}, where the above-mentioned problems were noted, and Eqs. (\ref{J})
and (\ref{L}) were rejected as unphysical. However, a Heidelberg group
\cite{heidelberg} took note of these equations and applied them to practical
problems on the grounds that they described correctly the electric and
magnetic fields of laser beams. Also, a Norwegian group \cite{selsto},
apparently without knowledge of Ref. \cite{hr79}, proposed these equations as
a way of introducing nondipole corrections into the study of laser-induced reactions.

While Eq. (\ref{J}) or (\ref{L}) is not acceptable for a properly formulated
theory, it is possible that qualitative information can be attainable from it.
It was used in Ref. \cite{hr63} to estimate the onset of magnetic effects,
which it established correctly.

\section{Solutions of Newton's equations are not necessarily physical}

Newtonian physics is based on forces, and electromagnetic forces are dependent
on electric and magnetic fields, as given by the Lorentz force expression%
\begin{equation}
\mathbf{F}=q\left(  \mathbf{E}+\frac{\mathbf{v}}{c}\mathbf{\times B}\right)  .
\label{N}%
\end{equation}
Hence, the reasoning applied to show the possibility of unphysical solutions
of the Maxwell equations applies as well to Newton's equations.

Alternative formulations of classical mechanics, such the Lagrangian,
Hamiltonian, Hamilton-Jacobi, ... are based on potentials, and hence they
convey more information than a force-based theory like Newton's mechanics.
This explains the common practice in mechanics textbooks to show that
potential-based formalisms imply the Newtonian formalism, but the reverse is
never shown.

\section{Practical consequences}

When approximations are employed in the study of a physical process, results
can be inefficient and possibly erroneous if basic symmetries are not
observed. An example from strong-field physics is the phenomenon known as
Above-Threshold Ionization (ATI), which refers to the observation \cite{ati}
that ionization by an intense laser beam can exhibit processes of photon
number in addition to, or in place of, the lowest allowed order predicted by
perturbation theory. AMO physics has experienced accurate and reliable results
from perturbation theory, and the observation of ATI came as a shock to the
AMO community. A recent assessment by prominent researchers \cite{cbc} of this
unexpected result can be paraphrased in abbreviated form as \textquotedblleft%
... multiphoton ionization experiments using intense infrared pulses found the
then-amazing result that an ionizing electron often absorbed substantially
more photons than the minimum needed for ionization. This puzzling behavior
led to the term ... ATI ... The problem was ultimately solved by computer
simulations and the semiclassical recollision model.\textquotedblright%
\ Citations to the relevant theory place the date for eventual understanding
of the 1979 experiment at 1993, a span of 14 years.

The important fact here is that both analytical and numerical studies employed
the dipole approximation, which has the effect of replacing the propagating
laser field by an oscillatory electric field. This loses the propagation
symmetry that is at the heart of strong-field processes described above.

From the point of view of propagating fields, the significant contribution of
many photon orders at high field intensities is obvious, and noted long before
the 1979 experiment. For example, in bound-bound transitions, there is the
1970 statement \cite{hr25} \textquotedblleft...as the intensity gets very high
... higher order processes become increasingly important.\textquotedblright%
\ For photon-multiphoton pair production in 1971 \cite{hr26}:
\textquotedblleft\ ... an extremely high-order process can ... dominate the
lowest order ...\textquotedblright. For interband transitions in band-gap
solids in 1977 \cite{hjhr}: \textquotedblleft...high-order processes can be
more probable than lower-order processes when the intensity is sufficiently
high.\textquotedblright\ The 1980 ionization paper \cite{hr80}, written before
the ATI\ experiment, describes ATI in detail. Other high-intensity phenomena,
such as channel closing and stabilization, are also discussed in the early
papers just cited.

\subsection{Nondipole corrections}

The difficulty of Eqs. (\ref{J}) or (\ref{L}) for the introduction of
nondipole corrections have been discussed above. A valuable laboratory project
would be to determine the limitations on such an approach.

A fully relativistic propagating strong-field theory is certainly applicable
for all nondipole, magnetic field, and relativistic studies. The construction
of such a theory was elaborated in Ref. \cite{hr90} for the Klein-Gordon case,
and implemented in detail in Ref. \cite{hrrel} for the Dirac case.

\subsection{Local Constant Field Approximation (LCFA)}

The LCFA is an example of how field-based criteria can differ from
potentials-based criteria. One justification of the LCFA follows from the
field-based observation that the two Lorentz invariants of plane-wave fields%

\begin{equation}
\mathbf{E}^{2}-\mathbf{B}^{2}=0,\quad\mathbf{E\cdot B}=0\mathbf{,} \label{O}%
\end{equation}
can be satisfied by constant crossed fields \cite{nikrit,ritus85}.

When viewed from the standpoint of potentials, the potentials that describe
constant crossed fields $\mathbf{E}_{0},\mathbf{B}_{0}$ are%
\begin{align}
\phi &  =-\mathbf{r\cdot E}_{0},\quad\mathbf{A}=-\frac{1}{2}\mathbf{r\times
B}_{0}\label{P}\\
\left\vert \mathbf{E}_{0}\right\vert  &  =\left\vert \mathbf{B}_{0}\right\vert
,\quad\mathbf{E}_{0}\perp\mathbf{B}_{0}. \label{Q}%
\end{align}
These potentials are unrelated to the $A^{\mu}\left(  \varphi\right)  $
requirement for propagating fields.

\subsection{Low frequency limit of a plane wave}

Plane waves are characterized by the fact that they propagate in vacuum at the
speed of light. This feature is independent of frequency. There is a line of
reasoning, adopted for many years in the strong-field community, that there
exists a zero-frequency limit of plane waves, and this limit is simply a
constant electric field. This is inferred from the dipole approximation, so
that there is no magnetic field present, distinguishing if from the LCFA.

There is no such thing as a zero frequency plane wave. Plane waves propagate
at the speed of light, independently of frequency. An example of a plane wave
phenomenon of extremely low frequency is the Schumann resonance
\cite{schumann}. This is a naturally occurring phenomenon in which powerful
lightning strikes generate extremely low frequency radio waves that resonate
in the cavity formed by the Earth's surface and the ionosphere. The lowest
mode of this cavity is $7.83Hz$, corresponding to a wavelength about equal to
the circumference of the Earth. On a laboratory scale, a plane wave with a
wavelength equal to the circumference of the Earth would appear to be a
constant field. Yet neither a constant crossed field nor a constant electric
field can spread its influence over the entire planet.

A pernicious consequence of the concept of a low frequency limit of a laser
field as being a constant electric field was its use as a criterion for
judging the worth of analytical approximations. For many years, a
zero-frequency limit equivalent to a constant electric field was regarded as a
feature of sufficient importance to reject any theory that did not possess
that property, See, for example, Ref. \cite{jkp}. This limit was regarded as
an adiabatic limit, and the qualitative stance was adopted that low frequency
fields should exhibit this adiabaticity. In actuality, the $\omega
\rightarrow0$ limit of plane waves is relativistic \cite{hr101}, not
adiabatic. It is the relativistic property of propagation at the speed of
light that distinguishes the Schumann resonance from a constant field phenomenon.

It is impossible to estimate the cost in valuable research resources of the
long-term application of the adiabaticity test as a basic criterion, but it is
undoubtedly considerable.

\section{Central role of $A^{\mu}$}

The basic properties of a propagating field can be described entirely by the
4-vector potential. This makes possible a covariant statement of those
properties, including the identity of the coupling constant of strong-field physics.

The 4-vector potential enters the description of propagating fields in the
three fundamental expressions:%
\begin{align}
\partial^{\mu}A_{\mu}  &  =0,\label{S}\\
k^{\mu}A_{\mu}  &  =0,\label{T}\\
z_{f}  &  \sim A^{\mu}A_{\mu}. \label{U}%
\end{align}
The first is the Lorenz condition, second is the transversality condition, and
the third enters into the definition of the strong-field coupling constant
$z_{f}$. The implications of Eq. (\ref{U}) seem to be little-known, but they
are perhaps the most direct expressions of the ascendancy of potentials over fields.

The Lorenz condition can be expanded into%
\begin{equation}
\partial^{\mu}A_{\mu}=\frac{\partial}{c\partial t}\phi-\mathbf{\nabla\cdot
A}=0. \label{V}%
\end{equation}
In the radiation (or Coulomb) gauge, where the scalar potential $\phi$ applies
only to longitudinal potentials, the Lorenz condition for the propagating
field reduces to $\mathbf{\nabla\cdot A}=0,$which is often used as the
identifying condition for the radiation gauge.

The expression (\ref{T}) is the covariant transversality condition. This is
readily shown to infer geometrical transversality: $\mathbf{k\cdot E}=0$ and
$\mathbf{k\cdot B}=0.$

The coupling constant of strong-field physics was identified
\cite{hrdiss,hr62} long ago. Strong-field physics as a separate discipline was
established \cite{hrepjd} as a consequence of the demonstration by Dyson
\cite{dyson} that standard QED\ (Quantum Electrodynamics) does not possess a
convergent perturbation expansion. This raised the question of the convergence
properties of an external-field theory, which represents a strong-field
situation where the number of photons present during an interaction is large.
The expansion parameter of standard QED is the fine-structure constant
$\alpha$. A convergence study of the external-field theory revealed the fact
that every appearance of $\alpha$ involved the same intensity-dependent
factor. That is, the expansion parameter is not $\alpha,$ but rather the
product of $\alpha$ with that factor. This product was labeled $z$ in the
original studies \cite{hrdiss,hr62}, since an expansion parameter must be
extended into the complex plane to find the singularities that limit
convergence, and $z$ is often used to label a complex number. In more recent
work $z$ was re-labeled $z_{f}$ to indicate that it is the intensity parameter
for free electrons as opposed to two new parameters $z$ (nonperturbative
intensity parameter) and $z_{1}$ (bound-state intensity parameter) that arise
when scalar potentials exist through interactions of the electron with binding
potentials in addition to the plane-wave field. See Section 1.3 in Ref.
\cite{hrrev} for further discussion.

In current terminology, the coupling constant is written%
\begin{equation}
z_{f}=2U_{p}/mc^{2}, \label{E1}%
\end{equation}
where $U_{p}$ is the ponderomotive energy, defined as%
\begin{equation}
U_{p}=\frac{e^{2}}{2mc^{2}}\left\langle \left\vert A^{\mu}A_{\mu}\right\vert
\right\rangle . \label{F1}%
\end{equation}
The angle brackets denote an average over a full cycle of the field, and the
absolute value is taken because $A^{\mu}$ is a spacelike 4-vector.

The quantity $z_{f}$ just identified as the coupling constant for strong laser
fields is already known as an intensity parameter for strong fields, but its
additional role as the coupling constant seems to have escaped general attention.

From Eq. (\ref{F1}) the ponderomotive energy and hence $z_{f}$ are Lorentz
invariants. If $z_{f}$ is to be a proper coupling constant it must also be
gauge-invariant, and this is not apparent in (\ref{F1}). However, when
$A^{\mu}$ describes a propagating field, then $U_{p}$ has been shown
\cite{hrjmo,hrup} to be gauge-invariant.

An objection is raised here to an intensity parameter that has found
acceptance in the relativistic strong-field literature. The quantity $A^{\mu
}A_{\mu}$ is rendered as $\left\vert \mathbf{E}^{2}\right\vert /\omega^{2},$
and then, since the square seems unnecessary, the parameter is commonly
written as proportional to $E/\omega$. The intent apparently is to introduce
the electric field in the belief that it is more fundamental than the 4-vector
potential. In addition to its inappropriate emphasis on the electric field
rather than on the 4-vector potential, this convention is objectionable in a
theory that is founded on covariant expressions. Lorentz-invariance is lost
because Lorentz transformation properties of the electric field $E$ are not
the same as for the frequency $\omega$. The quantity $z_{f}$ of Eq. (\ref{E1})
is Lorentz-invariant, gauge-invariant, and covariant.

It is not possible to express $A^{\mu}A_{\mu}$ in terms of fields. When a
quantity can be stated with fields it is always possible to convert it to
potentials because potentials are found from fields by differentiation; a
local procedure. To convert a quantity stated entirely with potentials, any
attempt to convert the expression to fields will fail because such a procedure
requires integration, which is nonlocal.

The $z_{f}$ parameter also occurs in the intensity-dependent mass-shell
condition for the electron in a strong field. The usual mass shell of QED is%
\begin{equation}
p^{\mu}p_{\mu}=\left(  mc\right)  ^{2}. \label{G1}%
\end{equation}
However, all the early studies of strong-field interactions
\cite{sengupta,hrdiss,hr62,nikrit0,bk,goldman} found the shifted-mass equation%
\begin{equation}
p^{\mu}p_{\mu}=\left(  mc\right)  ^{2}\left(  1+z_{f}\right)  . \label{H1}%
\end{equation}
Sarachik and Schappert find \cite{ss} that the altered mass shell expression
(\ref{H1}) also exists with classical strong fields.

A further implication of $z_{f}$ becomes clear when it is expressed in terms
of the photon density $\rho$. This expression has the form \cite{hrup}
\begin{equation}
z_{f}=\alpha\rho V,\label{I1}%
\end{equation}
with the fine structure constant $\alpha$ multiplied by the number of photons
contained in an effective interaction volume $V$. This volume is approximately
a cylinder of radius given by the electron Compton wavelength and a length
$\lambda$ given by the wavelength of the plane-wave field. The Compton
wavelength is the expected interaction length for a free electron, but
$\lambda$ is a macroscopic length in most laboratory applications. This is a
way to understand why $\omega\rightarrow0$ (or $\lambda\rightarrow\infty$)
leads to relativistic behavior and not adiabatic behavior.

The fundamental quantity $A^{\mu}A_{\mu}$ is expressed directly in terms of
the 4-vector potential $A^{\mu}$. There is no equivalent expression in terms
of the electric field. This simple compelling fact supports the primacy of
potentials over fields.

\section{The path ahead}

As laser intensities increase and as low-frequency capabilities improve, the
lessons contain herein are basic. In brief, one must consider the true
electromagnetic properties of very strong fields, including especially the
requirements that follow from the propagation property. The dipole
approximation, long a reliable feature of AMO physics, is not to be trusted,
and new criteria must be adopted that are consonant with relativistic
behavior. The penalty in waste of research resources that was mentioned in
connection with the explanation of ATI by a transverse-field method 14 years
earlier than in terms of the dipole approximation, and the fallacious
adiabaticity demand that also delayed progress by many years, is a caution
that also applies to the LCFA model. Guidance in research activities allow for
some investigation of the limits of applicability of the proposed nondipole
correction of Eq. (\ref{J}) or (\ref{L}), but always with the knowledge that
it lacks support as a reliable method.

Perhaps most important of all is the need to be aware that electromagnetic
potentials are the essential determinants of the nature of electromagnetic
phenomena, and that dependence on electric and magnetic fields carries
existential risks.

\end{document}